\documentclass[pra,10pt,showpacs]{revtex4}
\usepackage{graphicx,amsmath,amssymb}

\newcommand{\beq}{\begin{equation}}
\newcommand{\eeq}{\end{equation}}
\newcommand{\beqa}{\begin{eqnarray}}
\newcommand{\eeqa}{\end{eqnarray}}
\newcommand{\ket}[1]{| #1 \rangle}

\begin{document}


\title{Entanglement monotones for multi-qubit
states based on geometric invariant theory }

\author{Hoshang Heydari}
\email{hoshang@imit.kth.se} \affiliation{Institute of Quantum
Science, Nihon University, 1-8 Kanda-Surugadai, Chiyoda-ku, Tokyo
101- 8308, Japan}

\date{\today}

\begin{abstract}
We construct entanglement monotones for multi-qubit states based on
Pl\"{u}cker coordinate equations of Grassmann variety, which are
central notion in geometric invariant theory. As an illustrative
example, we in details investigate  entanglement monotones of a
three-qubit state.
\end{abstract}


\maketitle

\section{Introduction}
The geometry of entanglement is a very interesting subject as much
as quantification and classification of entanglement are
\cite{Dorje99,Moss,Miyake,Bern,Levay,Hosh3}.  It is possible to
derive geometrical measures of entanglement invariant under
Stochastic Local quantum Operation and Classical Communication
(SLOCC).  All homogeneous positive functions of pure states that are
invariant under determinant-one SLOCC operations are entanglement
monotones \cite{Verst03}. In this paper, we will derive
 entanglement monotones based on a  branch of the algebraic geometry
called geometric invariant theory. In particular, let  $G$ be a
group that acts on a set $A$, then the invariant theory is concerned
with the study of the fixed points $A^{G}$ and the orbits $A/G$
associated to this action. The geometric invariant theory deals with
the case where $G$ is an algebraic group, e.g., a special linear
group $SL(r,\mathbf{C})$, that acts on a variety $A$ via morphisms.
Thus, based on the geometric invariant theory, we can construct a
measure of entanglement that is invariant under action of
$SL(r,\mathbf{C})$ by construction. It has a well defined
geometrical structure called Grassmann variety or Grassmannian and
it is generated by a quadratic polynomials called the Pl\"{u}cker
coordinate equations. We will in detail discuss  our construction in
the following section. Recently, P\'{e}ter L\'{e}vay \cite{Levay1}
has constructed a class of multi-qubit entanglement monotones, which
was based on the construction  of C. Emary \cite{Emary}. His
construction based on bipartite partitions of the Hilbert space and
the invariants was expressed in terms of the Pl\"{u}cker coordinates
of the Grassmannian. However, we do have different approaches and
construction to solve the problem of quantifying multipartite
states, but some of the result on entanglement monotones for
multi-qubit states coincide. Now, let us start by denoting a
general, pure, composite quantum system with $m$ subsystems
$\mathcal{Q}=\mathcal{Q}^{p}_{m}(N_{1},N_{2},\ldots,N_{m})
=\mathcal{Q}_{1}\mathcal{Q}_{2}\cdots\mathcal{Q}_{m}$, consisting of
the pure state $
\ket{\Psi}=\sum^{N_{1}}_{k_{1}=1}\sum^{N_{2}}_{k_{2}=1}\cdots\sum^{N_{m}}_{k_{m}=1}
$ $\alpha_{k_{1},k_{2},\ldots,k_{m}} \ket{k_{1},k_{2},\ldots,k_{m}}
$ and corresponding the Hilbert space $
\mathcal{H}_{\mathcal{Q}}=\mathcal{H}_{\mathcal{Q}_{1}}\otimes
\mathcal{H}_{\mathcal{Q}_{2}}\otimes\cdots\otimes\mathcal{H}_{\mathcal{Q}_{m}}
$, where the dimension of the $j$th Hilbert space is given  by
$N_{j}=\dim(\mathcal{H}_{\mathcal{Q}_{j}})$. We are going to use
this notation throughout this paper. In particular, we denote a pure
two-qubit state by $\mathcal{Q}^{p}_{2}(2,2)$. Next, let
$\rho_{\mathcal{Q}}$ denote a density operator acting on
$\mathcal{H}_{\mathcal{Q}}$. The density operator
$\rho_{\mathcal{Q}}$ is said to be fully separable, which we will
denote by $\rho^{sep}_{\mathcal{Q}}$, with respect to the Hilbert
space decomposition, if it can  be written as $
\rho^{sep}_{\mathcal{Q}}=\sum^\mathrm{N}_{k=1}p_k
\bigotimes^m_{j=1}\rho^k_{\mathcal{Q}_{j}},~\sum^N_{k=1}p_{k}=1 $
 for some positive integer $\mathrm{N}$, where $p_{k}$ are positive real
numbers and $\rho^k_{\mathcal{Q}_{j}}$ denotes a density operator on
Hilbert space $\mathcal{H}_{\mathcal{Q}_{j}}$. If
$\rho^{p}_{\mathcal{Q}}$ represents a pure state, then the quantum
system is fully separable if $\rho^{p}_{\mathcal{Q}}$ can be written
as
$\rho^{sep}_{\mathcal{Q}}=\bigotimes^m_{j=1}\rho_{\mathcal{Q}_{j}}$,
where $\rho_{\mathcal{Q}_{j}}$ is the density operator on
$\mathcal{H}_{\mathcal{Q}_{j}}$. If a state is not separable, then
it is said to be an entangled state.

The general references for the complex projective space are
\cite{Griff78,Mum76}. So, let $\{f_{1},f_{2},\ldots,f_{q}\}$ be
continuous functions $\mathbf{K}^{n}\longrightarrow\mathbf{K}$,
where $\mathbf{K}$ is the  field of real $\mathbf{R}$ or complex
numbers $\mathbf{C}$. Then we define real (complex) space as the
set of simultaneous zeroes of the functions
\begin{eqnarray}
&&\mathcal{V}_{\mathbf{K}}(f_{1},f_{2},\ldots,f_{q})=\{(z_{1},z_{2},
\ldots,z_{n})\in\mathbf{K}^{n}:
f_{i}(z_{1},z_{2},\ldots,z_{n})=0~\forall~1\leq i\leq q\}.
\end{eqnarray}
These real (complex) spaces become  topological spaces by giving
them the induced topology from $\mathbf{K}^{n}$. Now, if all $f_{i}$
are polynomial functions in coordinate functions, then the real
(complex) space is called a real (complex) affine variety. A complex
projective space $\mathbf{CP}^{n}$ is defined to be the set of lines
through the origin in $\mathbf{C}^{n+1}$, that is,
$\mathbf{CP}^{n}=(\mathbf{C}^{n+1}-{0})/\sim$, where $\sim$ is an
equivalence relation define by $
(x_{1},\ldots,x_{n+1})\sim(y_{1},\ldots,y_{n+1})\Leftrightarrow\exists
\lambda\in \mathbf{C}-0$, such that  $\lambda x_{i}=y_{i} \forall
~0\leq i\leq n$. For $n=1$ we have a one dimensional complex
manifold $\mathbf{CP}^{1}$, which is a very important one, since as
a real manifold it is homeomorphic to the 2-sphere $\mathbf{S}^{2}$
e.g., the Bloch sphere. Moreover, every complex  compact manifold
can be embedded in some $\mathbf{CP}^{n}$. In particular, we can
embed a product of two projective spaces into the third one. Let
$\{f_{1},f_{2},\ldots,f_{q}\}$ be a set of homogeneous polynomials
in the coordinates $\{\alpha_{1},\alpha_{2},\ldots,\alpha_{n+1}\}$
of $\mathbf{C}^{n+1}$. Then the projective variety is defined to be
the subset
\begin{eqnarray}
&&\mathcal{V}(f_{1},f_{2},\ldots,f_{q})=\{[\alpha_{1},\ldots,\alpha_{n+1}]
\in\mathbf{CP}^{n}:
f_{i}(\alpha_{1},\ldots,\alpha_{n+1})=0~\forall~1\leq i\leq q\}.
\end{eqnarray}
We can view the complex affine variety
$\mathcal{V}_{\mathbf{C}}(f_{1},f_{2},\ldots,f_{q})\subset\mathbf{C}^{n+1}$
as a complex cone over the projective variety
$\mathcal{V}(f_{1},f_{2},\ldots,f_{q})$. We can also  view
$\mathbf{CP}^{n}$ as a quotient of the unit $2n+1$ sphere in
$\mathbf{C}^{n+1}$ under the action of $U(1)=\mathbf{S}^{1}$, that
is
$\mathbf{CP}^{n}=\mathbf{S}^{2n+1}/U(1)=\mathbf{S}^{2n+1}/\mathbf{S}^{1}$,
since every line in $\mathbf{C}^{n+1}$ intersects the unit sphere in
a circle.

\section{Grassmann variety}
In this section, we will define the Grassmann variety. However, the
standard reference on geometric invariant theory is \cite{Mum94}.
Let $\mathrm{Gr}(r,d)$ be the Grassmann variety of the
$r-1$-dimensional linear projective subspaces of
$\mathbf{CP}^{d-1}$. Now, we can embed $\mathrm{Gr}(r,d)$ into
$\mathbf{P}(\bigwedge^{r}(\mathbf{C}^{d}))=\mathbf{CP}^{\mathcal{N}}$,
$\mathcal{N}=\left(%
\begin{array}{c}
  d \\
  r \\
\end{array}%
\right)-1$, by using the Pl\"{u}cker map
$L\longrightarrow\bigwedge^{r}(L)$, where the exterior product
$\bigwedge^{r}(\mathbf{C}^{d})$ for $1\leq r \leq d$ is a subspace
of $\mathbf{C}^{N_{1}}\otimes\cdots\otimes\mathbf{C}^{N_{m}}$,
spanned by the anti-symmetric tensors. The Pl\"{u}cker coordinates
$P_{i_{1},i_{2},\ldots,i_{r}}, 1\leq i_{1}<\cdots<i_{r}\leq d$ are
the projective coordinates in this projective space. Next, let
$\mathbf{C}[\Lambda(r,d)]$ be a polynomial ring with the Pl\"{u}cker
coordinates $P_{J}$ indexed by elements of the set $\Lambda(r,d)$ of
ordered $r$-tuples in $\{1,2,\ldots d\}$ as its variables. Then the
image of the map $\kappa:\mathbf{C}[\Lambda(r,d)]\longrightarrow
\mathrm{Pol}(\mathrm{Mat}_{r,d})$, which assigns
$P_{i_{1}i_{2}\ldots i_{r}}$ the bracket polynomial
$[i_{1},i_{2},\ldots, i_{r}]$ (the bracket function on the
$\mathrm{Mat}_{r,d}$, whose values on a given matrix is equal to the
maximal minor formed by the columns from a set of $\{1,2,\ldots,
d\}$) is equal to the subring of the invariant of the polynomials.
Moreover, the kernel $\mathcal{I}_{r,d}$ of the map $\kappa$ is
equal to the homogeneous ideal of the Grassmann in its Pl\"{u}cker
embedding. Furthermore, the homogeneous ideal $\mathcal{I}_{r,d}$
defining $\mathrm{Gr}(r,d)$ in its Pl\"{u}cker embedding is
generated by the quadratic polynomials
\begin{eqnarray}
\mathcal{P}_{I,J}=\sum^{r+2}_{t=1}(-)^{t}P_{i_{1},\ldots,
i_{r-1},j_{t}}P_{j_{1}\ldots j_{t-1}j_{t+1},\ldots,j_{r+1}},
\end{eqnarray}
where $I=(i_{1}\ldots i_{r-1}), 1\leq i_{1}<\cdots<i_{r-1}< j_{i}$,
and $J=(j_{1},\ldots ,j_{r+1}), 1\leq j_{1}<\cdots<j_{r+1}\leq d$
are two increasing sequences of numbers from the set
$\{1,2,\ldots,d\}$. Note that the equations $\mathcal{P}_{I,J}=0$
define the Grassmannian $\mathrm{Gr}(r,d)$ are called the
Pl\"{u}cker coordinate equations. For example, for
$\mathrm{Gr}(2,d)$ and $n=2$, we have
\begin{eqnarray}
\mathcal{P}_{I,J}&=&\sum^{4}_{t=1}(-)^{t}P_{i_{1},j_{t}}P_{j_{1}\ldots
j_{t-1}j_{t+1},\ldots,j_{3}}\\\nonumber&=&
-P_{i_{1},j_{1}}P_{j_{2},j_{3}}+P_{i_{1},j_{2}}P_{j_{1},j_{3}}-P_{i_{1},j_{3}}P_{j_{1},j_{2}},
\end{eqnarray}
where $I=(i_{1})$, and $J=(j_{1},j_{2},j_{3})$. Note that, by its
construction, the Grassmannian $\mathrm{Gr}(2,d)$ is invariant under
$SL(2,\mathbf{C})$.

\section{Pl\"{u}cker coordinates and multipartite entanglement}
In this section, we will construct entanglement monotones based on
Pl\"{u}cker coordinate equations of the Grassmannian. Let us
consider a quantum system $\mathcal{Q}^{p}_{m}(2,2,\ldots,2)$ and
let
\begin{eqnarray}
\mathcal{E}_{I,J}(\mathrm{Mat}^{j}_{r,d})&=&\sum^{r+2}_{t=1}(
P^{i_{1},\ldots, i_{r-1},j_{t}}_{j}\overline{P}^{j}_{i_{1},\ldots,
i_{r-1},j_{t}}\\\nonumber&&+P^{j_{1}\ldots
j_{t-1}j_{t+1},\ldots,j_{r+1}}_{j}\overline{P}^{j}_{j_{1}\ldots
j_{t-1}j_{t+1},\ldots,j_{r+1}}),
\end{eqnarray}
where $I=(i_{1}\ldots i_{r-1}), 1\leq i_{1}<\cdots<i_{r-1}< j_{i}$,
and $J=(j_{1},\ldots ,j_{r+1}), 1\leq j_{1}<\cdots<j_{r+1}\leq d$
are two increasing sequences of numbers from the set
$\{1,2,\ldots,d\}$. For example, for $\mathrm{Gr}(2,d)$ and $r=2$,
that is invariant under $SL(2,\mathbf{C})$, we have
\begin{eqnarray}
\mathcal{E}_{I,J}(\mathrm{Mat}^{j}_{2,d})
&=&\sum^{4}_{t=1}(P^{i_{1},j_{t}}_{j}\overline{P}^{j}_{i_{1},j_{t}}+P^{j_{1}\ldots
j_{t-1}j_{t+1},\ldots,j_{3}}_{j}\overline{P}^{j}_{j_{1}\ldots
j_{t-1}j_{t+1},\ldots,j_{3}})\\\nonumber&&=
P^{i_{1},j_{1}}_{j}\overline{P}^{j}_{i_{1},j_{1}}
+P^{j_{2},j_{3}}_{j}\overline{P}^{j}_{j_{2},j_{3}}
+P^{i_{1},j_{2}}_{j}\overline{P}^{j}_{i_{1},j_{2}}\\\nonumber&&+
P^{j_{1},j_{3}}_{j}\overline{P}^{j}_{j_{1},j_{3}}
+P^{i_{1},j_{3}}_{j}\overline{P}^{j}_{i_{1},j_{3}}
+P^{j_{1},j_{2}}_{j}\overline{P}^{j}_{j_{1},j_{2}},
\end{eqnarray}
where $I=(i_{1})$, and $J=(j_{1},j_{2},j_{3})$.
 Now, we can write
the coefficient of a general multi-qubit state as follows
\begin{equation}
\begin{array}{c}
  \mathrm{Mat}^{1}_{2,d}=\left(%
\begin{array}{cccc}
  \alpha_{1,1,\ldots,1} & \alpha_{1,1,\ldots,2} & \ldots & \alpha_{1,2,\ldots,2}\\
  \alpha_{2,1,\ldots,1} & \alpha_{2,1,\ldots,2} & \ldots & \alpha_{2,2,\ldots,2}\\
\end{array}%
\right), \\
 \mathrm{Mat}^{2}_{2,d}=\left(%
\begin{array}{cccc}
  \alpha_{1,1,\ldots,1} & \alpha_{1,1,\ldots,2} & \ldots & \alpha_{2,1,\ldots,2}\\
  \alpha_{1,2,\ldots,1} & \alpha_{1,2,\ldots,2} & \ldots & \alpha_{2,2,\ldots,2}\\
\end{array}%
\right), \\
  \vdots \\
  \mathrm{Mat}^{m}_{2,d}=\left(%
\begin{array}{cccc}
  \alpha_{1,1,\ldots,1} & \alpha_{1,1,\ldots,1} & \ldots & \alpha_{2,2,\ldots,1}\\
  \alpha_{1,1,\ldots,2} & \alpha_{1,1,\ldots,2} & \ldots & \alpha_{2,2,\ldots,2}\\
\end{array}%
\right), \\
\end{array}
\end{equation}
where $d=2^{m-1}$ and $\mathrm{Mat}^{j}_{2,d}$, which we get by
permutation of $j$ for $1\leq j\leq m$. Moreover, we assume that the
sequences $I,J$ denote the columns of  the $\mathrm{Mat}^{j}_{2,d}$.
Then we can define entanglement monotones for the multi-qubit states
by
\begin{eqnarray}
\mathcal{E}(\mathcal{Q}^{p}_{m}(2,2,\ldots,2))&=&\left(\mathcal{N}\sum^{m}_{j=1}
\mathcal{E}_{I,J}(\mathrm{Mat}^{j}_{2,2^{m-1}})\right)^{1/2}.
\end{eqnarray}
As an example, let us consider the quantum system
$\mathcal{Q}^{p}_{3}(2,2,2)$. For such three-qubit states, if e.g.,
the subsystem $\mathcal{Q}_{1}$ is unentangled with the
$\mathcal{Q}_{2}\mathcal{Q}_{3}$ subsystems, then the separable set
of this state is generated by the six 2-by-2 subdeterminants of
\begin{equation}
\mathrm{Mat}^{1}_{2,4}=\left(%
\begin{array}{cccc}
  \alpha_{1,1,1} & \alpha_{1,1,2}&\alpha_{1,2,1}&\alpha_{1,2,2} \\
 \alpha_{2,1,1} & \alpha_{2,1,2}&\alpha_{2,2,1}&\alpha_{2,2,2} \\
\end{array}%
\right).
\end{equation}
$\mathrm{Mat}^{2}_{2,4}$ and $\mathrm{Mat}^{3}_{2,4}$ can be
obtained in similar way. Then the partial entanglement monotones for
$\mathrm{Mat}^{1}_{2,4}$ is given by
\begin{eqnarray}
\mathcal{E}_{I,J}(\mathrm{Mat}^{1}_{2,d})&=&
P^{i_{1},j_{1}}_{1}\overline{P}^{1}_{i_{1},j_{1}}
+P^{j_{2},j_{3}}_{1}\overline{P}^{1}_{j_{2},j_{3}}
+P^{i_{1},j_{2}}_{1}\overline{P}_{i_{1},j_{2}}+
P^{j_{1},j_{3}}_{1}\overline{P}^{1}_{j_{1},j_{3}}
\\\nonumber&&+P^{i_{1},j_{3}}_{1}\overline{P}^{1}_{i_{1},j_{3}}
+P^{j_{1},j_{2}}_{1}\overline{P}^{1}_{j_{1},j_{2}},
\end{eqnarray}
where the Pl\"{u}cker coordinates for $\mathrm{Mat}^{1}_{2,4}$ are
given by
\begin{eqnarray}
&&\nonumber
P^{1}_{1,2}=\alpha_{1,1,1}\alpha_{2,1,2}-\alpha_{1,1,2}\alpha_{2,1,1},~
P^{1}_{1,3}=\alpha_{1,1,1}\alpha_{2,2,1}-\alpha_{1,2,1}\alpha_{2,1,1},\\\nonumber&&
P^{1}_{1,4}=\alpha_{1,1,1}\alpha_{2,2,2}-\alpha_{1,2,2}\alpha_{2,1,1},
P^{1}_{2,3}=\alpha_{1,1,2}\alpha_{2,2,1}-\alpha_{1,2,1}\alpha_{2,1,2}
,\\\nonumber&&
P^{1}_{2,4}=\alpha_{1,1,2}\alpha_{2,2,2}-\alpha_{1,2,2}\alpha_{2,1,2},
P^{1}_{3,4}=\alpha_{1,2,1}\alpha_{2,2,2}-\alpha_{1,2,2}\alpha_{2,2,1}.
\end{eqnarray}
Thus, entanglement monotones for three-qubit states is given by
\begin{eqnarray}
\mathcal{E}(\mathcal{Q}^{p}_{m}(2,2,2))&=&\left(\mathcal{N}\sum^{3}_{j=1}\mathcal{E}_{I,J}(\mathrm{Mat}^{j}_{2,4})\right)^{1/2}.
\end{eqnarray}
Moreover, for matrices $\mathrm{Mat}^{j}_{2,4}$, we have $
\mathcal{P}^{j}_{I,J}=
-P^{j}_{1,2}P^{j}_{3,4}+P^{j}_{1,3}P^{j}_{2,4}-P^{j}_{1,4}P^{j}_{2,3}=0
$. 
For three-qubit states, this result coincides with construction of
the Segre variety \cite{Hosh3}. However, multi-qubit states needs
further investigation.  Now, as an example, let us consider the
state
$\ket{\Psi_{W}}=\alpha_{1,1,2}\ket{1,1,2}+\alpha_{1,2,1}\ket{1,2,1}
+\alpha_{2,1,1}\ket{2,1,1}$. Then we have
\begin{eqnarray}\nonumber
    \mathcal{C}(\mathcal{Q}_{3}(2,2,2))&=&(2\mathcal{N}
[ \left|\alpha_{1,2,1}\alpha_{2,1,1} \right|^{2}+\left |
\alpha_{1,1,2}\alpha_{2,1,1} \right|^{2}+\left
|\alpha_{1,1,2}\alpha_{1,2,1}
 \right|^{2}])^{1/2}.
\end{eqnarray}
In particular, for
    $\alpha_{1,1,2}=\alpha_{1,2,1}=\alpha_{2,1,1}=\frac{1}{\sqrt{3}}$, we get
    $\mathcal{C}(\mathcal{Q}_{3}(2,2,2))=
    (\frac{2}{3}\mathcal{N})^{1/2}$.
\section{Hyperdeterminant and Pl\"{u}cker coordinate equations}
In this section, we will review some result of the construction of
entanglement measure based on the hyperdeterminant for three-qubit
states and relation between the Pl\"{u}cker coordinate equations and
the hyperdeterminant. We also discuss a generalization of this
construction. The hyperdeterminant of the elements of
$\mathbf{C}^{N_{1}}\otimes\mathbf{C}^{N_{2}}\otimes\cdots\otimes\mathbf{C}^{N_{m}}$
was introduced by Gelfand, Kapranov, and Zelevinsky in \cite{GKZ}.
They proved that the dual variety of Segre product
$\mathbf{CP}^{N_{1}-1}\times\mathbf{CP}^{N_{2}-1}\times\cdots\times\mathbf{CP}^{N_{m}-1}$
is a hypersurface if and only if $N_{j}\leq\sum_{i\neq j}N_{i}$ for
$j=1,2,\ldots,m$. Whenever the dual variety is a hypersurface its
equation is called the hyperdeterminant of the format $N_{1}\times
N_{2}\times\cdots\times N_{m}$ and denoted by $\mathrm{Det}$. The
hyperdeterminant is a homogeneous polynomial function over
$\mathbf{C}^{N_{1}}\otimes\mathbf{C}^{N_{2}}\otimes\cdots\otimes\mathbf{C}^{N_{m}}$,
so that the condition $\mathrm{Det} A\neq0$ is meaningful for
$\mathcal{A}\in \mathbf{CP}^{N_{1}N_{2}\cdots N_{m}-1}$. Moreover,
the hyperdeterminant $\mathrm{Det}$ is $SL(N_{1},\mathbf{C})\times
SL(N_{2},\mathbf{C}))\otimes\cdots\otimes
SL(N_{m},\mathbf{C})$-invariant. For example, for $m=2$ we have
$\mathrm{Det}\mathcal{A}=\alpha_{1,1}\alpha_{2,2}-\alpha_{1,2}\alpha_{2,1}$
and for $m=3$, we have
\begin{eqnarray}\label{HH}
\mathrm{Det}\mathcal{A}&=& \alpha^{2}_{1,1,1}\alpha^{2}_{2,2,2}
+\alpha^{2}_{1,1,2}\alpha^{2}_{2,2,1}+\alpha^{2}_{1,2,1}\alpha^{2}_{2,1,2}+
\alpha^{2}_{2,1,1}\alpha^{2}_{1,2,2}\\\nonumber&&-
2(\alpha_{1,1,1}\alpha_{1,1,2}\alpha_{2,2,1}\alpha_{2,2,2}
+\alpha_{1,1,1}\alpha_{1,2,1}\alpha_{2,1,2}\alpha_{2,2,2}\\\nonumber&&+
\alpha_{1,1,1}\alpha_{2,1,1}\alpha_{1,2,2}\alpha_{2,2,2}
+\alpha_{1,1,2}\alpha_{1,2,1}\alpha_{2,1,2}\alpha_{2,2,1}\\\nonumber&&+
\alpha_{1,1,2}\alpha_{2,1,1}\alpha_{1,2,2}\alpha_{2,2,1}
+\alpha_{1,2,1}\alpha_{2,1,1}\alpha_{1,2,2}\alpha_{2,1,2})\\\nonumber&&+
4(\alpha_{1,1,1}\alpha_{2,2,1}\alpha_{2,1,2}\alpha_{1,2,2}
+\alpha_{2,2,2}\alpha_{2,1,1}\alpha_{1,2,1}\alpha_{1,1,2}).
\end{eqnarray}
Now, let us introduce the Diophantine function. For any sequence of
number $\gamma_{1},\gamma_{2},\gamma_{3},\gamma_{4}$, and
$\delta_{1},\delta_{2},\delta_{3},\delta_{4}$ we have
\begin{eqnarray}\label{HH1}
&&\mathcal{P}(\gamma_{1},\gamma_{2},\gamma_{3},\gamma_{4},
\delta_{1},\delta_{2},\delta_{3},\delta_{4})
=(\gamma_{1}\delta_{1}+\gamma_{2}\delta_{2}-\gamma_{3}\delta_{3}-\gamma_{4}\delta_{4})^{2}
-4(\gamma_{1}\gamma_{2}+\delta_{3}\delta_{4})(\gamma_{3}\gamma_{4}+\delta_{1}\delta_{2}).
\end{eqnarray}
This equation is equal to the hyperdeterminant
$\mathrm{Det}\mathcal{A}$. For the quantum system
$\mathcal{Q}^{p}_{3}(2,2,2)$ we have
\begin{eqnarray}\label{HH1}
\mathrm{Det}\mathcal{A}&=&\mathcal{P}(-\alpha_{1,1,1},\alpha_{2,2,1},
\alpha_{2,1,2},\alpha_{1,2,2}
,-\alpha_{2,2,2},\alpha_{1,1,2},\alpha_{1,2,1},\alpha_{2,1,1}).
\end{eqnarray}
 For the quantum system
$\mathcal{Q}^{p}_{3}(2,2,2,2)$, one can wonder if it would be
possible to find a generalization of the polynomial  $\mathcal{P}$
such that the hyperdeterminant $\mathrm{Det}\mathcal{A}$ could be
give by $\mathcal{P}$. We can also construct the hyperdeterminant in
terms of the Pl\"{u}cker coordinates. For example, P. L\'{e}vay
\cite{Levay} has constructed such Pl\"{u}cker coordinates for
three-qubit states as follows. Let
$\alpha^{p},\beta^{q},p,q=1,2,3,4$ be two-four component vectors
defined as
$\alpha_{1,k_{1},k_{2}}=\frac{1}{\sqrt{2}}\alpha_{p}\Sigma^{p,k_{1},k_{2}}$,
and
$\alpha_{2,k_{1},k_{2}}=\frac{1}{\sqrt{2}}\beta_{p}\Sigma^{p,k_{1},k_{2}}$,
where $\Sigma^{s}=-i\sigma_{s}, s=1,2,3$, and $\Sigma^{4}=I_{2}$,
where $\sigma_{p}$ are Pauli matrices. Then the hyperdeterminant is
given by $\mathrm{Det}\mathcal{A}=2P_{p,q}P^{p,q}$, where the
Pl\"{u}cker coordinates are given by
$P_{p,q}=\alpha_{p}\beta_{q}-\alpha_{q}\beta_{p}$. Thus, this
construction can be at least extended into multi-qubit states
following our definition of the Pl\"{u}cker coordinates for the
multi-qubit states and further extending the definition of two
$2^{m-1}$ components vectors
$\alpha^{p},\beta^{q},p,q=1,2,\ldots,2^{m-1}$. For further progress
in this direction see Ref. \cite{Levay1}.
\section{Conclusion}
In this paper, we have constructed entanglement monotones for
multipartite states based on the Grassmannian $\mathrm{Gr}(r,d)$,
which was  defined in terms of the Pl\"{u}cker coordinate equations.
In particular, we have given an explicit expression for entanglement
monotones for multi-qubit states. Moreover, we have investigated
entanglement monotones  for three-qubit state as an illustrative
example.
\begin{flushleft}
\textbf{Acknowledgments:}  The author acknowledges useful comments
from Jan Bogdanski and Jonas S\"{o}derholm and useful discussions
with P\'{e}ter L\'{e}vay. This work was supported by the Wenner-Gren
Foundations.
\end{flushleft}


\end{document}